\documentclass[aps,prb,twocolumn,groupedaddress]{revtex4}

\usepackage[dvips]{color}
\usepackage{graphicx,color}
\begin{document}


\title{Probing Non-Abelian Statistics in $ \nu=12/5 $ Quantum Hall State}


\author{K.T. Law}
\affiliation{Department of Physics, Brown University, Providence, Rhode Island 02912, USA}



\begin{abstract}

The tunneling current and shot noise of the current between two Fractional Quantum Hall (FQH) edges in the $ \nu=12/5 $ FQH  state in electronic Mach-Zehnder interferometer are studied. It is shown that the tunneling current and shot noise can be used to probe the existence of $k=3 $ parafermion statistics in the $ \nu=12/5 $ FQH state. More specifically, the dependence of the current on the Aharonov-Bohm flux in the Read-Rezayi state is asymmetric under the change of the sign of the applied voltage. This property is absent in the Abelian Laughlin states. Moreover the Fano factor can exceed 12.7 electron charges in the $ \nu=12/5 $ FQH  state . This number well exceeds the maximum possible Fano factor in all Laughlin states and the $ \nu=5/2 $ Moore-Read state which was shown previously to be $ e $ and $ 3.2 e $ respectively.

\end{abstract}

\pacs{73.43.Jn, 71.10.Pm, 73.43.Fj}

\maketitle

\section{\label{Intro} Introduction}

Particles other than bosons and fermions can exist in two dimensions. One possibility is that when one particle makes a circle around another particle, the total many-particle wave function acquires a non-trivial phase factor $ e^{i\phi}$  where $ \phi $ can be a real number, not constrained to be 0 or $ 2\pi $. Those particles are called Abelian anyons. \cite{LM, Wil} A more exotic situation can happen when the state of the system is described by a multi-component state vector as the positions of the particles are specified. The actions of braiding a particle around another are represented by unitary matrices acting on the state vector. If the braiding matrices do not commute with each other, the particles under study are called non-Abelian anyons. \cite{GMS,Fr}

Both Abelian and non-Abelian anyons are proposed to exist in Fractional Quantum Hall (FQH) systems. Arovas et al.\cite{ASW} show that the fractionally charged quasiparticles in Laughlin states obey fractional statistics with $ \phi=2\pi\nu $ where $ \nu $  is the filling factor. In the paper by Moore and Read,\cite{MR} it was argued that quasiparticles in FQH systems with filling factor $ \nu=5/2 $  possibly obey non-Abelian statistics. Later, Read and Rezayi suggested that quasiparticles in the more recently observed $ \nu=12/5 $  FQH state \cite{Exp1,Exp2,Exp3} may also obey non-Abelian statistics but with an even richer structure,\cite{RR} which can support universal topological quantum computing. \cite{FKLW}

Several theoretical proposals have been made to probe Abelian and non-Abelian anyons in FQH systems. Using an elegant two point contact interferometer with an antidot in the middle of a quantum hall bar to probe Abelian anyons in Laughlin states was initially proposed in Ref. \onlinecite{CFKSW}. Later, the same idea was extended to probing non-Abelian quasiparticles.\cite{FNTW,DFN,SH,BKS,GSS,BSS,CS} Other proposals are also available,\cite{SDM,KLVF,Kan,IGS} but there are no experimental realizations of those ideas so far. Recently an electronic Mach-Zehnder interferometer (MZI) in the integer quantum Hall regime, whose schematic diagram shown in Fig. \ref{f1}, has been fabricated at the Weizmann Institute.\cite{CSHMS} Unlike the Fabry-Perot interferometer proposed in Ref. \onlinecite{CFKSW} in which the interference pattern can be destroyed by fluctuations of the number of quasiparticles trapped inside the interferometer, the MZI interferometer is free from this limitation, as long as the fluctuations are sufficiently slow compared with the quasiparticle tunneling rate. We believe that a similar device in the FQH regime can be realized with higher magnetic fields and will provide a practical way to probe the existence of non-Abelian anyons.

In this paper, we suggest that the tunneling current and shot noise of the current between two FQH edges in MZI geometry with two quantum point contacts (QPCs) can be used to probe the existence of non-Abelian anyons in $ \nu=12/5 $ FQH state. We show that: i) The tunneling current can be reduced to a sinusoidal form $ I(\Gamma_1, \Gamma_2) = I_{0}(\Gamma_1, \Gamma_2) + I_{\Phi}(\Gamma_1, \Gamma_2)\cos(2\pi\Phi_a +const) $ when $ \Gamma_2 \ll \Gamma_1 $, where $ \Gamma_1 $ and $ \Gamma_2 $ are the tunneling amplitudes at the QPCs shown in Fig. \ref{f1} and $ \Phi_a $ denotes the magnetic flux enclosed by the two FQH edges. There is a scaling relation between $I_0 $ and $ I_{\Phi} $. The scaling exponent $ b $, defined as $  I_{\Phi}(\Gamma_1, \Gamma_2) \sim [ I_{0}(\Gamma_1 , \Gamma_2)-I_{0}(\Gamma_1 , 0)]^b $, equals to $ 5/2 $ in the $ \nu=12/5 $ Read-Rezayi state. This value is the same as the one in $ \nu=1/5 $ Laughlin state but different from $ b=2 $ in the $ \nu=5/2 $ Moore-Read state.\cite{LFG,FK} ii) The flux dependence of the tunneling current is asymmetric under the change of the sign of the applied voltage. iii) The Fano factor, which is defined as the ratio between the shot noise and the tunneling current, can be as large as 12.7 in units of one electron charge in the $ \nu=12/5 $ state. As it was previously shown in Ref. \onlinecite{FGKLS}, the maximum Fano factor is one electron charge and 3.2 electron charge for the Laughlin states and the $ \nu=5/2 $ Moore-Read state respectively. The last two properties, the asymmetric I-V curve and larger than one electron charge Fano factor, are direct consequences of the non-trivial fusion rules and braiding rules of the quasiholes. Their observations would provide experimental evidence for the existence of non-Abelian statistics in the $ \nu=12/5 $ FQH state.

Calculating the tunneling current and shot noise in the MZI geometry is more complicated than in the simple Fabry-Perot geometry.\cite{CFKSW} This is due to the fact that the number of quasiparticles trapped inside the interferometer changes when a quasiparticle tunnels from one edge to the other. As a result, the statistical phase due to the quasiparticles enclosed by the interference paths changes. Hence, the tunneling probabilities of the tunneling quasiparticles changes accordingly after every tunneling event.\cite{LFG,FK,FGKLS} However, we show that in the $ \nu=12/5 $ FQH state, quasiparticles trapped inside the interferometer can fuse together to form only ten non-equivalent classes of states (or superselection sectors) which are characterized by their electric and  topological charges. Tunneling of a quasihole from one edge to another edge changes one state into another.  The transition rates between the ten non-equivalent classes depend on the fusion rules and braiding rules of the quasiparticles as well as other experimental parameters. We will calculated the tunneling current and the corresponding shot noise in sections V and VI.  For the purpose of illustration, some of the relevant results\cite{LFG,FK,FGKLS} concerning the $ \nu=5/2 $ FQH state are reproduced throughout this paper.

This paper is organized as follows.  In section II, the structure of electronic MZI is explained. In section III, we work out the fusion rules and braiding rules for the quasiparticles in the $\nu=5/2 $ and $ \nu=12/5 $ FQH states. In section IV, we calculate the transition rates between the non-equivalent classes in the MZI. In section V and VI, the tunneling current and shot noise is calculated. Section VII presents the conclusion.

\section{Electronic Mach-Zehnder interferometer}

In this section, the structure of an electronic MZI will be explained. We will also see how the non-trivial statistical phase $ \phi_s $ can affect the tunneling current and noise dramatically. 

\begin{figure}
\includegraphics[width=3.2in]{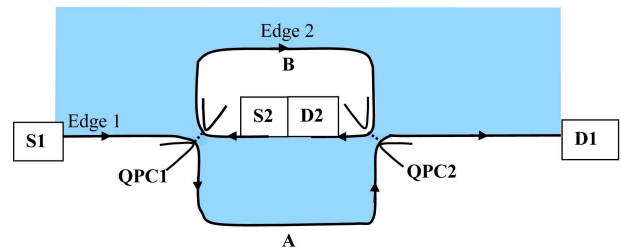}
\caption{\label{f1} Schematic picture of the Mach-Zehnder interferometer. S and D denote sources and drains. Arrows indicate the propagation direction of the chiral edge modes. Quasiparticles tunnel between edges 1 and 2.}
\end{figure}

A schematic diagram of an electronic MZI is depicted in Fig. \ref{f1}. S1, S2, D1, D2 denote the sources and drains of the corresponding FQH edges 1 and 2. The arrows on the edges indicate the edge mode propagation directions. A and B are two points on the edges. Quasiparticles on the edges are allowed to tunnel from one edge to the other through two QPCs denoted by QPC1 and QPC2 respectively. As the bulk excitations are gaped, the low energy physics of the MZI is determined by the edges. Hence, the Hamiltonian can be written as
\begin{equation}
\hat{H}=\hat{H}_{edge} + [(\Gamma_1\hat{T}_{1}+\Gamma_2\hat{T}_{2})+H.c.],  \label{hamiltonian}
\end{equation}
where $ \hat{H}_{edge} $ denotes the Hamiltonian for the two edges and  $ \hat{T}_1 $ and $ \hat{T}_2 $ are tunneling operators which transfer a quasihole from edge 1 to edge 2 at QPC1 and QPC2 respectively. 

A voltage difference $ V $ between S1 and S2 will result in a tunneling current from one edge to the other, say, from edge 1 to edge 2. For a quasiparticle to arrive at D2 from S1, there are two possible tunneling paths S1-QPC1-A-QPC2-D2 and S1-QPC1-B-QPC2-D2. In the integer Quantum Hall regime when the tunneling quasiparticles are electrons, the transition rate depends on the tunneling amplitudes $ \Gamma_1 $ and $ \Gamma_2 $ as well as the Aharonov-Bohm phase $ \phi_{AB}={2\pi \Phi_{a}}/{\Phi_{0}} $, where $ \Phi_a $ is the magnetic flux enclosed by the two tunneling paths. For small $ \Gamma_1 $ and $ \Gamma_2 $, the transition rate can be written as\cite{LFG}
\begin{equation}
p=r_{0}\{(|\Gamma_1|^2 + |\Gamma_2|^2) + 2u|\Gamma_1^*\Gamma_2|\cos[2\pi{\Phi_a}/{\Phi_0}+\delta]\} ,  \label{tr1}
\end{equation}
where $ r_{0} $ and $ u $ are functions of temperature $ T $, the applied voltage $ V $ and the interferometer size $ L $ whose exact form can be calculated by using the Hamiltonian in Eq. (\ref{hamiltonian}), $ \delta=\arg(\Gamma_1^*\Gamma_2)$. The tunneling current in the integer Quantum Hall regime is simply $ I_{I} = e p $.\cite{LFG} The zero frequency shot noise can be written as $ S_{I} = e I_{I} $.\cite{FGKLS} The Fano factor $ S_I / I_I $ is independent of the applied magnetic flux and equals $1$ in the units of an electron charge.

In the FQH regime, the tunneling quasiparticles are anyons with fractional electric charge $ qe$, where $q$ is a fractional number. In addition to the Aharonov-Bohm phase $ \phi_{AB}= {2\pi q\Phi_{a}}/{\Phi_{0}} $, a tunneling quasiparticle experience a statistical phase $ \phi_s $ originated from other quasiparticles enclosed by the tunneling paths.

In the Laughlin states with $ \nu=1/m $, $ \phi_s = n 2\pi \nu $ where $ n $ is the number of quasiparticles inside the interferometer. The transition rate can be written as:
\begin{equation}
p_{n}=r_{0}\{(|\Gamma_1|^2 + |\Gamma_2|^2) + 2u|\Gamma_1^*\Gamma_2|\cos[2\pi q {\Phi_a}/{\Phi_0}+ \phi_{s}+\delta]\}.   \label{tr2}
\end{equation}
It is important to note that the transition rate depends on $ n $ mode $ m $. At zero temperature, there is a simple way to calculate the tunneling current. Let us assume that there are $ n=km $ quasiparticles inside the interferometer initially. Then, the transition rate is $ p_0 $ and the average time to transfer a quasiparticle from S1 to D2 is $ t_0 = 1/p_0 $. After one quasiparticle tunneling, $ n $ is increased by one, the transition rate becomes $ p_1 $ and $ t_1 = 1/p_1 $. After $ m $ tunneling events, the transition rate returns to the initial value. The total time needed to transfer $ m $ quasiparticles is $ \bar{t}=\sum\limits_{i=0}^{m-1} t_i $. Hence, in terms of the transition rates, the tunneling current in the $ \nu=1/m $ Laughlin state is $ I_{1/m}=e/(\sum\limits_{n=0}^{n=m-1}1/p_{n}) $. This zero temperature result together with the finite temperature ones can be derived rigorously using the chiral Luttinger liquid theory of the edge states.\cite{LFG} Moreover, one can also show that the zero frequency shot noise is $ S_{1/m} = e^2 ({\sum\limits_{n=0}^{n=m-1}1/{p^2_{n}}})/{(\sum\limits_{n=0}^{n=m-1}1/p_{n})^3} $.\cite{FGKLS} The Fano factor is flux dependent since the tunneling rates $ p_n$ depend on the applied magnetic flux. Finally, we have $ S_{1/m}/eI_{1/m}= ({\sum\limits_{n=0}^{n=m-1}1/{p^2_{n}}})/{(\sum\limits_{n=0}^{n=m-1}1/p_{n})^2} \le 1 $.\cite{FGKLS}

For the non-Abelian states, the situation is more complicated. A quasiparticle in the non-Abelian states is characterized by its electric charge and its topological charge. The statistical phase $ \phi_s $ depends on the number of quasiparticles inside the interferometer as well as the topological charges of the tunneling quasiparticle and of the quasiparticles inside the interferometer. The statistical phase $ \phi_s $ in the cases of the $ \nu = 5/2 $ Moore-Read state and the $ \nu=12/5 $ Read-Rezayi state will be calculated in the section III. Moreover, the result of fusing two topological charges together may not be unique. For instance, the topological charge of a quasihole in the $5/2$ state is $ \sigma $. According to the fusion rules, two $ \sigma $ fields can fuse together and the resulting field can be $ I $ or $ \psi $ with equal probability $1/2 $. The factor $ 1/2 $ modifies the transition rates. In the $ \nu=12/5 $ FQH state, the situation is slightly more complicated. The transition rates will be studied in detail in section IV.

\section{Statistical phase}

Generalizing the idea of Moore and Read,\cite{MR} Read and Rezayi pointed out that  the $ \nu=5/2 $ and $ \nu=12/5 $ states can be described by the $ k=2 $ and $ k=3 $ parafermion conformal field theories respectively.\cite{RR} More specifically, a quasiparticle operator can be written as $ \Psi_{q.p}=\Phi_{m}^{l} V_{\alpha} $ where the parafermion field $ \Phi_{m}^{l} $ describes the topological charge of the quasiparticle and the vertex operator $ V_{a}=:e^{ia\phi_c}: $ describes its electric charge. The braiding properties between quasiparticles depend on both of the quantum numbers. In this section, we review the fusion rules and braiding rules of the parafermion theories and the chiral boson theory. The statistical phase $ \phi_s $ acquired when a quasiparticle makes a full circle around another in the non-Abelian states is calculated.

\subsection{ Parafermion and chiral boson theories }

The parafermion conformal field theory\cite{ZF,GQ} has central charge $ c=\frac{2k-2}{k+2}$ in the Virasoro algebra. The primary fields in the theory are labeled as $ \Phi_{l}^{l} $ and have conformal dimension $ h_{l}=\frac{l(k-l)}{2k(k+2)} $ where $ l=0,1,...,k-1 $. Each $ \Phi_{l}^{l} $ generates a series of fields $ \Phi_{m}^{l} $ with conformal dimensions
\begin{equation}
\begin{array}{c}
h_{m}^{l}= h_{l}+ \frac{(l-m)(l+m)}{4k},\qquad  \textrm{for}\, -l \leq m < l,    \\
h_{m}^{l}= h_{l}+ \frac{(m-l)(2k-l-m)}{4k},\qquad \textrm{for}\, l \leq m \leq 2k-l.             \label{cd}
\end{array}
\end{equation}
The conformal fields are subject to the constraints $ l+m \equiv 0 \ (\textrm{mod}2) $ and $ \Phi_{m}^{l}=\Phi_{k+m}^{k-l}=\Phi_{m+2k}^{l} $. The fusion rules and the operator product expansions for the conformal fields are given in Refs. \onlinecite{ZF} and \onlinecite{GQ} and can be written as:
\begin{equation}
\Phi_{m}^{l}\times\Phi_{m'}^{l'}= \sum\limits_{n=|l-l'|}^{\min(l+l',2k-l-l')}\Phi_{m+m'}^{n} \label{fr1}.
\end{equation}
\begin{equation}
\Phi_{m}^{l}(z)\Phi_{m'}^{l'}(0)=\sum\limits_{n}C_{mm'}^{ll'n}z^{\Delta h}\Phi_{m+m'}^{n}(0) \label{pope}
\end{equation}
where the expansion coefficients $ C_{mm'}^{ll'n} $ are constants and $ \Delta h = h_{m+m'}^{n} - h_{m}^{l} - h_{m'}^{l'} $. The exponent $ \Delta h $ in Eq. (\ref{pope}) is important as it gives the statistical phase, $ 2\pi\Delta h $, acquired when a conformal field $ \Phi_{m}^{l} $ makes a full circle around another conformal field $ \Phi_{m'}^{l'} $ when the result of fusing these two fields is $ \Phi_{m+m'}^{n} $.

On the other hand, a free chiral bosonic field is governed by the action 
\begin{equation}
S=-\frac{1}{4\pi}\int dxdt [\partial_t\phi_c\partial_x\phi_c + ({\partial_x\phi_c})^2 ].
\end{equation}
The vertex operator $ V_{a}=:e^{ia\phi_c}: $ has conformal dimension $ \frac{a^2}{2} $.\cite{DMS} The operator product expansions between vertex operators have the form
\begin{equation}
V_{a}(z)V_{b}(0)=C_{ab}z^{ab}V_{a+b}(0)  \label{vope}
\end{equation}
so that the fusion rule between vertex operators is
\begin{equation}
V_{a}\times V_{b}=V_{a+b}.   \label{fr2}
\end{equation}

\subsection{ $\nu=5/2$ Moore-Read state}

The $ \nu=5/2 $ Moore-Read state can be described by the $ k=2 $ parafermion theory. According to IIIA, we can easily see that $ \Phi^{0}_{0} $, $ \Phi^{0}_{2} $ and $ \Phi^{1}_{1} $ are the only three independent fields in the theory. Following the notations in Ref. \onlinecite{RR}, we label the fields as $ I $, $ \psi $ and $ \sigma $ respectively. The conformal dimension of the three fields are $ 0 $, $ \frac{1}{2} $ and $ \frac{1}{16} $ respectively according to Eq. (\ref{cd}). Following Eq. (\ref{fr1}), the fusion rules are: $ \psi \times \psi =1$, $ \psi \times \sigma = \sigma $ and $ \sigma \times \sigma = 1 + \psi$.

One of the important observations of Moore and Read is that the ground state trial wavefunction of the $ \nu=5/2 $ FQH state can be expressed as the correlation function of operators of the form $ \psi :e^{\sqrt{2}\phi_c}: $,
\begin{equation}
\begin{array}{l}
\Psi_{5/2}=Pf(\frac{1}{z_{i}-z_{j}})\prod_{i > j}(z_{i}-z_{j})^2 =\\ 
<\psi(z_1),\psi(z_2)\cdots \psi(z_N)e^{i\sqrt{2}\phi{(z_1)}}e^{i\sqrt{2}\phi{(z_2)}}\cdots e^{i\sqrt{2}\phi{(z_N)}}\Phi_{bg}>,  \label{correlator}
\end{array}
\end{equation}
where $Pf$ is the Pfaffian and $ \Phi_{bg}= e^{-i\int d^2z \sqrt{2}\rho_{0}\phi{(z)}}$ is the background charge operator with $\rho_{0} $ denotes the charge density. Evidently, the electron operator can be identified as $ \Psi_{5/2,el.} = \psi :e^{\sqrt{2}\phi_c}: $. Moreover, one can show that the states with quasiholes can be obtained by inserting operators $  \sigma :e^{\frac{1}{2\sqrt{2}}\phi_c}: $ into the correlator in Eq. (\ref{correlator}). As a result, the quasihole operator can be written as $ \Psi_{5/2,q.h.}= \sigma :e^{\frac{1}{2\sqrt{2}}\phi_c}: $. In other words, a quasihole carries electric charge $ \frac{e}{4}$ and topological charge $ \sigma $. With the identification of the quasiparticle operators with the conformal fields, we may calculate the statistical phases $ \phi_s $ acquired when a quasihole makes a full circle around another excitation with electric charge $ n\frac{e}{4} $ and topological charge $ \alpha $. In the $ \nu = 5/2 $ state, $\alpha $ can take three values: $ I$, $\psi $ and $\sigma$.

The statistical phase  can be written as the sum of two contributions:
\begin{equation}
\phi_s = n\frac{\pi}{4}+ \phi_{\sigma \alpha}^{\beta}.  \label{sp}
\end{equation}
The first term on the right hand side of Eq. (\ref{sp}) originates from the vertex operators. It can be calculated from substituting $ a=\frac{1}{2\sqrt{2}}$ , $ b=\frac{n}{2\sqrt{2}} $ and $ z=e^{2\pi} $ into Eq. (\ref{vope}). The second term $ \phi_{\sigma \alpha}^{\beta} $ originates from the parafermion fields where $ \alpha $ denotes the topological charge of the other excitation  and $ \beta $ denotes the fusion result of $ \sigma $ and $ \alpha $. According to Eqs. (\ref{fr1}) and (\ref{pope}), if $ \alpha = I $ , $ \phi_{\sigma I}^{\sigma}=0 $; if $ \alpha = \psi$, $ \phi_{\sigma \psi}^{\sigma}=\pi $; however, if $ \alpha = \sigma $, there are two possibilities, $ \phi_{\sigma \sigma}^{I}=\frac{-\pi}{4} $ and $ \phi_{\sigma \sigma}^{\psi}=\frac{3\pi}{4}$. These results are consistent with the ones derived by using the algebraic theory of anyons in Ref. \onlinecite{FK}.

\subsection{$\nu=12/5$ Read-Rezayi state}

Read and Rezayi proposed that  the $ \nu=12/5 $ FQH state can be described by the $k=3 $ parafermion theory. According to IIIA, one can find six nonequivalent conformal fields in this theory. Following the notations in Refs. \onlinecite{RR} and \onlinecite{CS}, we define the six nonequivalent fields as $ \Phi_{0}^{0}=I $, $ \Phi_{2l}^{0}= \psi_{l} $, $ \Phi_{l}^{l}=\sigma_{l} $ and $ \Phi_{0}^{2}= \epsilon $, where $l=1,2$. The fusion rules which are relevant to our calculations can be written as:
\begin{equation}
\begin{array}{cc}
I \times \sigma_l = \sigma_l, & \psi_{l} \times \sigma_{l} = \epsilon,  \\
\psi_{3-l} \times \sigma_{l} = \sigma_{3-l}, & \epsilon \times \sigma_{l} = \psi_{3-l} + \sigma_{l},  \\
\sigma_{l} \times \sigma_{l} = \psi_{l} + \sigma_{3-l}, & \sigma_{3-l} \times \sigma_{l} = I +\epsilon.
\end{array}
\end{equation}
The ground state trial wave function of the FQH state can be obtained by the correlation function of electron operators with the form $ \Psi_{12/5,el.}= \psi_{1}:e^{i\sqrt{\frac{5}{3}}\phi_c}: $ and the quasihole operator can be written as $ \Psi_{12/5,q.h.} =\sigma_{1}:e^{i\sqrt{\frac{1}{15}}\phi_c}: $. Evidently, a quasihole carries $ \frac{e}{5} $ electric charge and topological charge $ \sigma_1 $. Similar with the case of the $ \nu=5/2 $ state, the statistical phase of a quasihole making a full circle around an excitation with electric charge $ n \frac{e}{5} $ and topological charge $ \alpha $ can be written as the sum of two contributions $ \phi_s = n \frac{2\pi}{15} + \phi_{\sigma_1, \alpha}^{\beta} $. The first contribution can be obtained by substituting $ a = \frac{1}{\sqrt{15}} $, $ b = n \frac{2\pi}{\sqrt{15}} $ and $ z=e^{i 2 \pi} $ into Eq. (\ref{vope}). The non-Abelian contribution $ \phi_{\sigma_1, \alpha}^{\beta} = 2\pi(h_{\beta}-h_{\alpha}-h_{\sigma_1}) $ can be calculated from the conformal dimensions of the fields. The results are listed in Table I.

\begin{table}
\caption{\label{T1} Statistical phase originating from topological charges in the $ \nu=12/5 $ state}
\begin{ruledtabular}
\begin{tabular}{|c|c|c|c|c|c|c|c|c|}
 $\phi_{\sigma_1\alpha}^\beta $ & $ \phi_{\sigma_1\epsilon}^{\psi_2} $ & $ \phi_{\sigma_1\epsilon}^{\sigma_1} $ & $ \phi_{\sigma_1{\psi_1}}^\epsilon $ & $ \phi_{\sigma_1{\psi_2}}^{\sigma_2} $ & $ \phi_{\sigma_1{\sigma_1}}^{\psi_1} $ & $ \phi_{\sigma_1{\sigma_1}}^{\sigma_2} $ & $ \phi_{\sigma_1{\sigma_2}}^{I} $ & $ \phi_{\sigma_1{\sigma_2}}^{\epsilon} $ \\
\hline
Phase & $ 2\pi\frac{1}{5} $ & $ 2\pi\frac{3}{5} $ & $ 2\pi\frac{2}{3} $ & $ 2\pi\frac{1}{3} $ & $ 2\pi\frac{8}{15} $ & $ 2\pi\frac{14}{15} $ & $ 2\pi\frac{13}{15} $ & $ 2\pi\frac{4}{15} $ \\
\end{tabular}
\end{ruledtabular}
\end{table}

\section{Transition rates \label{TR}}

\begin{figure}
\includegraphics[width=2.8in]{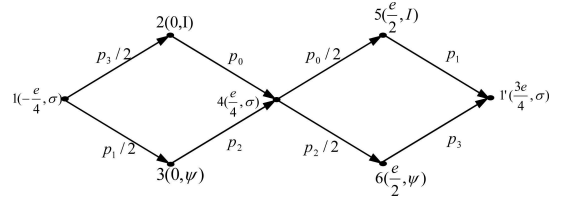}
\caption{\label{5half} The numbers from 1 to 6 labels the six non-equivalent classes of the composite particle enclosed by the interferometer in the $ \nu=5/2 $ Moore-Read state. The quantum number inside the bracket denote the electric charge and the topological charge of the state respectively. The arrows indicate the transition direction at zero temperature. $p_l$ denote the transition rates and the factors of 1/2 are due to the fusion probabilities.}
\end{figure}

In MZI geometry, the quasiholes inside the area enclosed by QPC1-A-QPC2-B-QPC1 can be seen as a composite particle with electric charge $ nqe $, and topological charge $ \alpha $  from the point of view of a tunneling quasihole on the FQH edges. In the $ \nu=5/2 $ state, for n quasiholes, the electric charge is $ ne/4 $ and the topological charge $ \alpha $ may take value $ I $, $ \psi $ or $ \sigma $, as a result of fusing $ n $ $ \sigma $ fields. When $ n $ is odd, the topological charge is $ \sigma $. When $ n $ is even, the topological charge can be $ I $ or $ \psi $. We can obtain six non-equivalent classes of states by fusing $ n $ quasiholes. The six classes are: 1($ \frac{-e}{4}, \sigma $), 2($ 0, I $), 3($ 0, \psi $), 4($ \frac{e}{4}, \sigma $), 5($ \frac{e}{2}, I $) and 6($\frac{e}{2}, \psi $). The numbers in the bracket represent the electric charge (mod $e$) and the topological charge respectively. The six classes are represented as vertexes in Fig. \ref{5half}. State ($\frac{3e}{4}, \sigma$) is identified with state ($\frac{-e}{4},\sigma$) since they differ from each other by an electron. The arrows represent the transition directions between states at zero temperature. The transition rates are labeled as $ p_k $ where
\begin{equation}
p_k=r_{0}\{(|\Gamma_1|^2 + |\Gamma_2|^2) + 2u|\Gamma_1^*\Gamma_2|\cos[2\pi{\Phi_a}/{4\Phi_0}+ k\pi/2+\delta]\}.  \label{tp2}
\end{equation}

These results can be obtained from Eq. (\ref{tr2}) with $ q=\frac{1}{4} $ and the results for $ \phi_s $ in section IIIB. The transition rates from $ 1 \to 2 $, $ 1 \to 3 $, $ 4 \to 5 $ and $ 4 \to 6 $ are modified by a factor of $ 1/2 $ which is called the fusion probability. Heuristically, if the quasiparticle inside the interferometer has topological charge $ \sigma $, the fusion result of the quasiparticle with a quasihole is not unique according to the fusion rule $ \sigma \times \sigma = 1 + \psi $. There are equal probabilities that the resulting field is $ I $ or $ \psi $. One may follow the arguments in Refs. \onlinecite{Kit,FGKLS} to calculate the fusion probabilities $ p_{\sigma\sigma}^{I} $ and $ p_{\sigma\sigma}^{\psi} $.

\begin{figure}
\includegraphics[width=1.5in]{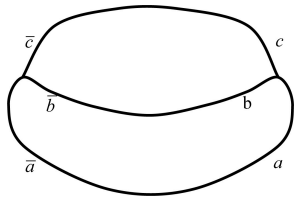}
\caption{\label{fa} Two pairs of particles $ \bar{a}a $ and $ \bar{b}b $ are created out of the vacuum and particles $ a(\bar{a}) $ and $ b (\bar{b}) $ are fused together to result in particle $ c (\bar{c})$.}
\end{figure}

In the context of algebraic theory of anyons, suppose two particle-antiparticle pairs $ a \bar{a} $ and $ b \bar{b} $ are created out of the vacuum, if $ a $ and $ b $ are fused together, what is the probability $ p_{ab}^{c} $ that the resulting particle is c? This process is depicted in Fig. \ref{fa}, the amplitude of the process is shown to be $ \sqrt{\frac{d_{c}}{d_a d_{b}}} $, where $ d_{\alpha} $ denotes the quantum dimension of a particle with topological charge $ \alpha $. Hence, the fusion probability is\cite{Kit,Pre,FGKLS}
\begin{equation}
p_{ab}^{c}= \sum_{i}\frac{d_{c}}{d_a d_{b}}= N_{ab}^{c}\frac{d_{c}}{d_a d_{b}} ,  \label{fp}
\end{equation} 
where $ N_{ab}^{c} $ is called the fusion multiplicity which gives the number of ways that $ a $ and $ b $ can be fused together to result in $ c $.

In the $ \nu=5/2 $ state, $ d_I = d_{\psi} = 1 $ and $ d_{\sigma} = \sqrt{2} $. $ N_{ab}^{c} $ can be zero or $ 1 $ according to the fusion rules. As a result, we have $ p_{\sigma \sigma}^{I} = p_{\sigma \sigma}^{\psi}=1/2 $.

\begin{figure}
\includegraphics[width=3in]{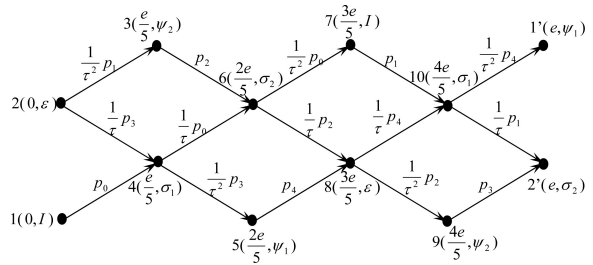}
\caption{\label{f4} The numbers from 1 to 10 labels the ten non-equivalent classes of the composite particle enclosed by the interferometer in the $\nu=12/5 $ FQH state . The quantum number inside the bracket denote the electric charge and the topological charge of the state respectively. The arrows indicate the transition direction at zero temperature. $ p_{\sigma \alpha_a}^{\beta_b} P_{a \to b}^{+} $ denote the forward transition rates.}
\end{figure}

In the $ \nu=12/5 $ FQH state, the fusion results of the quasiparticles inside the interferometer can be classified into ten non-equivalent classes. The ten classes are represented by the vertexes in Fig. \ref{f4}, from state $1$ to state $10$. The quantum numbers inside the bracket denote the electric charge and the topological charge of the state. State $ 1 (2) $ and state $ 1' (2') $ are in the same class because they differ from each other by  an electron which can be labeled as $ ( -e, \psi_1 )$. The transition rates between the states are denoted by $ p_{l} $ where
\begin{equation}
p_{l}=r_{0}\{(|\Gamma_1|^2 + |\Gamma_2|^2) + 2u|\Gamma_1^*\Gamma_2|\cos[2\pi({\Phi_a}/{\Phi_0}+l)/5+\delta]\}. \label{tp}
\end{equation}
Similar with the case in the $ \nu=5/2 $ FQH state, $ p_{l} $ can be obtained from Eq. (\ref{tr2}) and the results of the statistical phase $ \phi_s = n \frac{2\pi}{15} + \phi_{\sigma_1, \alpha}^{\beta} $ from section III. The transition rates are modified by the corresponding fusion probability. Quasiparticles with topological charge $ I, \psi_1 $ and $ \psi_2 $ have quantum dimension $ 1 $ and quasiparticles with topological charge $ \sigma_1, \sigma_2 $ and $ \epsilon $ have quantum dimension $ \tau=2cos(\pi/5) $. Hence, from Eq. (\ref{fp}), the fusion probabilities can take three values $ 1, \frac{1}{\tau}$ and $ \frac{1}{\tau^2} $. For example, when a quasihole with electric charge $ e/5 $ and topological charge $ \sigma_1 $ fuses with a quasiparticle in state $ 1 ( 0, I )$, the resulting state has electric charge $ e/5 $ and topological charge $ \sigma_1 $ with probability 1. On the other hand, if a quasihole fuse with a quasiparticle in state $ 2 (0,\epsilon) $, the resulting state can be $ 3(e/5, \Psi_2 ) $ or $ 4(e/5,\sigma_1) $ with probability $ \frac{1}{\tau^2} $ and $ \frac{1}{\tau} $ respectively.

At zero temperature, when the voltage difference $ V $ between S1 and D2 is positive, only transitions indicated by the arrows in Fig. \ref{f4} can happen. For convenience, we denote this forward transition (the transition along the directions of the arrows) rate from state $ a $ to state $ b $ as $ P_{a \to b}^{+} $. The total forward transition rate, defined as $ P_{a \to b}^{+} $ modified by the corresponding fusion probability, is denoted as $ R_{a \to b}^{+} = p_{\sigma_1 \alpha_a}^{\beta_b} P_{a \to b}^{+} $, where $\alpha_a $ and $ \beta_b $ denotes the topological charge of the state before and after the fusion with a quasihole.

At finite temperature, quasiholes can tunnel from  edge 2 to edge 1 such that transitions in Fig. \ref{f4} can occur in directions both alone and against the directions of the arrows. A backward tunneling event can be regarded as the following physical process: a quasihole-quasiparticle pair is created out of the vacuum near the tunneling point contact, the quasihole tunnels to edge 1 while the quasiparticle with charge $ -e/5 $ and topological charge $ \sigma_2 $ tunnels into edge 2 and fuses with the quasiholes enclosed by the interference paths. As a result, the backward transition rate $ R_{a\to b}^{-} = p_{\sigma_2 \alpha_a}^{\beta_b}P_{a \to b}^{-}$ where $ P_{a \to b}^{-}=P_{b \to a}^{+}e^{{-eV}/{5k_{B}T}} $ is obtained from the detailed balance condition.

\section{Zero temperature current}

After classifying the states and obtaining the transition rates between them in section \ref{TR}, we are ready to calculate the tunneling current between the two FQH edges with the help of Figs. \ref{5half} and \ref{f4}. In this section, we assume that the condition $ k_B T \ll qeV $ is satisfied such that quasiparticle tunneling can happen only from one edge to the other. 

In the $ \nu=5/2 $ FQH state, there are four possible ways to transfer one electron from S1 to D2. The four possible ways are denoted as four different paths in Fig. \ref{5half}. If we take state $ 1(0,I) $ as the starting point, the four paths are 1-2-4-5-1', 1-2-4-6-1',1-3-4-5-1' and 1-3-4-6-1' respectively. The average time it takes to transfer an electron by path 1-2-4-5-1' is $ t_{1}= \frac{1}{p_{3}} + \frac{1}{p_{0}}+\frac{1}{p_{0}}+\frac{1}{p_{1}} $. The tunneling probabilities $ p_k $ are given in Eq. (\ref{tp2}). The probability to take path 1 is $ q_1 = \frac{p_3}{p_3 + p_1} \frac{p_0}{p_0 + p_2} $. Similarly, we can denote the average time taken to transfer an electron through path $i$ by $ t_i $ and the corresponding probability by $ q_i $. The average time to transfer an electron from S1 to D2 is $ \bar{t}_{5/2}=\sum\limits_{1}^{4} t_{i}q_{i} $ and the current is $ I_{5/2}=e/{\bar{t}_{5/2}} $. See Ref. \onlinecite{FGKLS} for a full derivation. The explicit expression of the tunneling current in terms of the tunneling probabilities is:\cite{FGKLS}
\begin{equation}
I_{5/2}(V)=\frac{e}{ \frac{1}{p_1 + p_3} (2+\frac{p_3}{p_0}+\frac{p_1}{p_2})+ \frac{1}{p_0 +p_2}(2+ \frac{p_0}{p_1}+ \frac{p_2}{p_3})}.
\end{equation}
It is important to note that when we change the sign of the applied voltage, the tunneling current can be calculated in the same way as above. However, the corresponding diagram in Fig. \ref{5half} is modified in two ways. First, the directions of the arrows are reversed. More importantly, the fusion probabilities are modified from $ 1/2 (1) $ to $ 1 (1/2)$. Hence, the resulting current is different from the original one besides a change in the sign of the current. In terms of the tunneling probabilities, we have:
\begin{equation}
I_{5/2}(-V)=\frac{e}{ \frac{1}{p_1 + p_3} (2+\frac{p_1}{p_0}+\frac{p_3}{p_2})+ \frac{1}{p_0 +p_2}(2+ \frac{p_2}{p_1}+ \frac{p_0}{p_3})}.
\end{equation}

This observation is significant because it is a result of the non-unity fusion probability and this property is absent in the Laughlin states. The experimental observation of the asymmetric I-V curve provides evidence for the existence of the non-Abelian excitations in the $\nu=5/2 $ FQH state.\cite{FK} As shown below, the I-V curve in the $ \nu=12/5 $ state is also asymmetric.

The diagram shown in Fig. \ref{f4} for the $ \nu=12/5 $ FQH state is considerably more complicated than the ones for the Laughlin states and $ \nu=5/2$ state. In the rest of this section, we will calculate the zero temperature current with the kinetic equation approach.

The transitions between the 10 non-equivalent states labeled in Fig. \ref{f4} are governed by the kinetic equations for the system:\cite{LFG,FGKLS}
\begin{equation}
\frac{df_i(t)}{dt}=\sum\limits_{j=1}^{10}[{-f_i(t)(R_{i \to j}^+ + R_{i \to j}^-) + f_j(t)(R_{j \to i}^+ + R_{j \to i }^-)} ], \label{ke}
\end{equation}
where $ f_i(t) $  is the probability of the composite particle inside the interferometer to be found in state $ i $ at time $ t $. In this notation, the tunneling current has the expression:
\begin{equation}
I=e^*{\sum\limits_{i=1}^{10}}f_i{\sum\limits_{j=1}^{10}}{(R_{i \to j}^+ - R_{i \to j}^-)}, \label{ftcurrent}
\end{equation}
where $ f_i $ are the steady state solutions of Eq. (\ref{ke}) and $ e^{*} $ equals $ e/5 $. At zero temperature, the backward transition rates $ R_{i \to j}^- $ equal zero. In this case, the steady state solutions $ f_i $ for the kinetic equations can be found easily in the following way:

First, we define $ f'_{k}=f_{k}\sum\limits_{j=1}^{10}{R_{k \to j}^+ }$, $ \tilde{R}_{i-1 \to i}=\frac{R_{i-1 \to i}^{+}}{R_{i-1 \to i}^{+}+R_{i-1 \to i+1}^{+}} $ and $ \tilde{R}_{i-1 \to i+1}=\frac{R_{i-1 \to i+1}^{+}}{R_{i-1 \to i}^{+}+R_{i-1 \to i+1}^{+}} $, where  $ i={1,3,5,7,9} $ and we have used the notation $ n=n+10 $. With the steady state condition $ \frac{df_i(t)}{dt}=0 $, we can rewrite the kinetic equations into a set of matrix equations:
\begin{equation}
\left(
\begin{array}{c}
f'_{i} \\ f'_{i+1}
\end{array}
\right)=\left(
\begin{array}{cc}
0 & \tilde{R}_{i-1 \to i} \\
1 & \tilde{R}_{i-1 \to i+1}
\end{array}
\right){\left(
\begin{array}{c}
f'_{i-2} \\ f'_{i-1}
\end{array}
\right)} .  \label{me1}
\end{equation}
From Eq. (\ref{me1}), we can see that $ f'_{i}+f'_{i+1}= C $ where $ C $ is independent of $ i $ because $ \tilde{R}_{i-1 \to i} + \tilde{R}_{i-1 \to i+1} = 1 $. As a result, the tunneling current can be written as $ I=5 e^* C $. Denoting the 2 by 2 matrix in Eq. (\ref{me1}) by $ M_{i-1} $, we obtain a self-consistent matrix equation for $ f'_{1} $ and $ f'_{2} $ which reads:
\begin{equation}
\left(
\begin{array}{c}
f'_{1} \\ f'_{2}
\end{array}
\right)
= M_{10}M_{8}M_{6}M_{4}M_{2}
\left(
\begin{array}{c}
f'_{1} \\ f'_{2}
\end{array}
\right)  . \label{me2}
\end{equation}

The solution of Eq. (\ref{me2}) is $ f'_{1}=N[1-({\tilde{R}_{2 \to 4}}{\tilde{R}_{4 \to 5}}+{\tilde{R}_{2 \to 4}}{\tilde{R}_{4 \to 6}}{\tilde{R}_{6 \to 8}}+{\tilde{R}_{2 \to 3}}{\tilde{R}_{6 \to 8}}){\tilde{R}_{8 \to 9}}]{\tilde{R}_{10 \to 1}}=Nf''_{1}$ and $ f'_{2}=N[1-(1-({\tilde{R}_{4 \to 5}}+{\tilde{R}_{4 \to 6}}{\tilde{R}_{6 \to 8}}){\tilde{R}_{8 \to 9}}){\tilde{R}_{10 \to 1}}]=Nf''_{2} $ where $ N $ is a normalization factor. $ f''_{1} $ ($ f''_{2}$) can be interpreted as the transition probability to state 1 (2) from state 2 (1) after 5 tunneling events. Using Eq. (\ref{me1}), we can generate $ f'_{i}=Nf''_{i} $ from $ f'_{1} $ and $ f'_{2} $. From the solutions of Eq. (\ref{me2}), one can easily show that $ f'_{1}+f'_{2}= f'_{i}+f'_{i+1}= N(1+ {\tilde{R}_{2 \to 3}}{\tilde{R}_{4 \to 5}}{\tilde{R}_{6 \to 7}}{\tilde{R}_{8 \to 9}}{\tilde{R}_{10 \to 1}})=C $. With the normalization condition $ \sum\limits_{l=1}^{10}f_{l} =1 $, we conclude that 
\begin{equation}
N = \frac{1}{\sum\limits_{l=0}^{4}[f''_{2l+1} \frac{1}{R_{2l+1 \to 2l+4}^+} + f''_{2l+2}\frac{1}{R_{2l+2 \to 2l+3}^+ + R_{2l+2 \to 2l+4}^+}]}.     \label{nor}
\end{equation}
As a result, in terms of the forward transition rates, the tunneling current can be expressed formally as:
\begin{equation}
I_{12/5} = \frac{5 e^* (1+ {\tilde{R}_{2 \to 3}}{\tilde{R}_{4 \to 5}}{\tilde{R}_{6 \to 7}}{\tilde{R}_{8 \to 9}}{\tilde{R}_{10 \to 1}})}{\sum\limits_{l=0}^{4}[f''_{2l+1} \frac{1}{R_{2l+1 \to 2l+4}^+} + f''_{2l+2}\frac{1}{R_{2l+2 \to 2l+3}^+ + R_{2l+2 \to 2l+4}^+}]}.    \label{current}
\end{equation}

The form of the functional dependence of the current on the transition rates in Eq. (\ref{current}) does not depend on the specific values of the transition rates. However, exploring the symmetry in the transitions rates in Fig. \ref{f4}, we have 
\begin{equation}
\left(
\begin{array}{c}
f'_{1+2k}(\Phi_a) \\ f'_{2+2k}(\Phi_a)
\end{array}
\right)=
\left(
\begin{array}{c}
f'_{1}(\Phi_a+{2k'\Phi_{0}}) \\ f'_{2}(\Phi_a+2k' \Phi_{0})
\end{array}
\right)
\end{equation}
where $ 2k'= 2k \ mod \ 5 $ and $ k $ runs from $ 0 $ to $4$. As a result, the expression of the tunneling can be expressed in a more manageable form.
\begin{equation}
 I_{12/5} = \frac{5 e^* (1+ {\tilde{R}_{2 \to 3}}{\tilde{R}_{4 \to 5}}{\tilde{R}_{6 \to 7}}{\tilde{R}_{8 \to 9}}{\tilde{R}_{10 \to 1}})}{ \sum\limits_{k=0}^{5}[f''_{1}\frac{1}{R_{1 \to 4}^{+}}({\Phi_a + 2k'\Phi_{0}})+ f''_{2}\frac{1}{R_{2 \to 3}^{+}+ R_{2 \to 4}^{+}}({\Phi_a + 2k'\Phi_{0}})]}   \label{current2}
\end{equation}
Eq. (\ref{current2}) can be further simplified as
\begin{equation}
I_{12/5} = 5 e^{*} \frac{r_{0}(|\Gamma_1|^2 + |\Gamma_2|^2)[\alpha + \beta \cos(2\pi\Phi_a + 5\delta)]}{\gamma + \kappa \cos(2\pi\Phi_a +5\delta) + \xi \sin(2\pi\Phi_a +5\delta)}   \label{current3}
\end{equation}
where $\alpha$, $\beta$, $\gamma$, $\kappa$, $\xi$ are functions of $ R=\frac{u|\Gamma_1^*\Gamma_2|}{|\Gamma_1|^2 + |\Gamma_2|^2} $ as well as the applied voltage but independent of the applied magnetic flux. Their exact algebraic expressions are lengthy and will not be shown here. The important point is that $ \alpha, \gamma \sim 1 + ... $ and $ \beta,\kappa,\xi \sim R^5 + ... $ where $ ... $ denotes the higher order terms in $ R $. When $ R \ll 1 $, for example, when $ \Gamma_2 \ll \Gamma_1 $, the tunneling current can be reduced to a sinusoidal form $ I = I_{0} + I_{\Phi}\cos(2\pi\Phi_a +const) $. In this limit, one can show that 
\begin{equation}
 I_{\Phi}(\Gamma_1, \Gamma_2) \sim [ I_{0}(\Gamma_1 , \Gamma_2)-I_{0}(\Gamma_1 , 0)]^b  \label{ep}
\end{equation}
with $ b= 5/2 $. The exponent $ b $ in the $ k=3 $ Read-Rezayi state is the same as the one in the $ \nu=1/5 $ Laughlin states\cite{LFG} but different from the one in the Moore-Read state in which $ b=2 $.\cite{FK}

If we change the sign of the voltage difference between S1 and S2 such that quasiholes propagate from S2 to D1, a similar diagram as the one in Fig. \ref{f4} with different transition rates is obtained, resulting in a different functional dependence of the tunneling current on the applied magnetic flux. More specifically, the coefficient $ \xi $ of $ sin(2\pi\Phi_a +5\delta) $ in Eq. (\ref{current3}) will change its sign to $ -\xi $ besides an overall change in the sign of the current.

In order to work out the diagram with $ V $ changed to $ -V $, we can imagine the following physical process: a quasihole-quasiparticle pair is created out of the vacuum near S2, the quasihole with electric charge $ e/5 $  and topological charge $ \sigma_1 $ will tunnel to D1 through two possible paths, S2-QPC1-B-QPC2-D1 and S2-QPC1-A-QPC2-D1. The quasiparticle with charge $ -e/5 $ and topological charge $ \sigma_2 $ will fuse with the composite particle inside the area enclosed by QPC1-A-QPC2-B-QPC1. Accordingly, we can obtain a new diagram similar to Fig. \ref{f4} and calculate the current with the procedures mentioned above. The new diagram can be obtained by changing the direction of the arrows in Fig. \ref{f4} with the fusion probabilities modified from $ \frac{1}{\tau^2} $(1) to $ 1 $($ \frac{1}{\tau^2} $).

We assume that the interferometer size $ L $ and the excitation velocity satisfy the condition $ eVL/5h\nu \ll 1 $ such that the function $ u $ in Eq. (\ref{tp}) can be set to 1. Moreover, we absorb the function $ r_0 $ into the tunneling amplitudes $ \Gamma_1 $ and $ \Gamma_2 $. The flux dependence of the tunneling current $I(\Gamma_{1}, \Gamma_{2})$ in Eq. (\ref{current2}) is plotted in Fig. \ref{current} with $ \Gamma_{1}=1 $ in the appropriate units in all four curves. For the three solid curves from the one with the largest amplitude to the one with the smallest amplitude, $ \Gamma_2$ equals $1, 0.9 $ and $ 0.8 $ respectively. The dashed curve depicts the magnitude of the current when $ V $ is changed to $ -V $ with $ \Gamma_2=1 $. We set the arbitrary phase $ \delta=0 $ in the calculation. 

\begin{figure}
\includegraphics[width=3.2in]{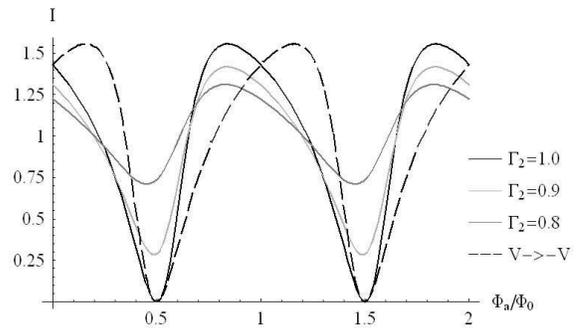}
\caption{\label{current} Shows the flux dependent of $I(\Gamma_{1}, \Gamma_{2})$. $ \Gamma_1=1 $ for all the four curves. For the three solid curves, from the one with the largest amplitude to the one with the smallest amplitude, $ \Gamma_2 $ equals to $ 1, 0.9 $ and $0.8 $ respectively. The dashed curve depicts the magnitude of the current when $ V $ is changed to $ -V $ with $ \Gamma_2=1 $.}
\end{figure}

From Eq. ({\ref{current2}}) we see that the current is a periodic function of the magnetic flux with period $ \Phi_{0} $, this agrees with the Byers-Yang theorem.\cite{BY} The minimums of the current occur at $ \Phi_{a}/\Phi_{0}= 1/2 \, \bmod \, 1 $ when $ \delta=0 $. This can be explained by the fact that whenever $ \Phi_{a}/\Phi_{0}= 1/2 \, \bmod  \, 1  $, one of the tunneling probabilities $ p_{l} $ in Eq. ({\ref{tp}}) can be much smaller than 1 given that $ u \approx 1 $. For example, if $ p_{2} $ is very small, the system will be ``trapped'' at state $3(\frac{e}{5},\psi_2) $ for a long time before any tunneling event can happen. This results in a minimum in the tunneling current.

\section{Zero temperature shot noise}

It was pointed out in Ref. \onlinecite{FGKLS} that measuring the Fano factor $ \tilde{e} $, which is defined as the ratio between the shot noise and the tunneling current, can be an effective way to probe non-Abelian statistics. It was shown that the maximum Fano factor in the Laughlin state and the Moore-Read state is $ e $ and $ 3.2e $ respectively where $ e $ is an electron charge. It is shown below that the maximum Fano factor in the $ k=3 $ Read-Rezayi state can be as large as $ 12.7e $. 

Shot noise is defined as the Fourier transform of the current-current correlation function,
\begin{equation}
S(\omega)= {1\over 2}\int_{-\infty}^{+\infty}<\hat{I}(0)\hat{I}(t)+ \hat{I}(t)\hat{I}(0)> \exp(i\omega t)dt.
\end{equation}
In this paper, we are interested in the low frequency limit of the shot noise. In this limit, $ S $ can be written as\cite{FGKLS} 
\begin{equation}
S=<\delta Q^2(t)>/t \label{sn}
\end{equation}
where $ \delta Q(t)$ is the fluctuation of the charge $ Q(t) $ transmitted during a period of time $ t $ when the measurement time $ t $ is fixed.

In this section, we use the generating function method developed in Ref. \onlinecite{FGKLS} to calculate the zero temperature shot noise and the Fano factor. Without loss of generality, we can assume that the fusion result of the quasiparticle inside the interferometer at time $ t=0 $  is in state 1 or state 2 as labeled in Fig. \ref{f4}. Let us define $ P_{k+5n,i}(t) $ as the probability that $ k+5n $ quasiparticles have transferred from S1 to D2 at time $ t $ and the resulting composite particle is in state $ i $. Here, we denote $ k=[\frac{i-1}{2}] $ where $ [y] $ is the largest integer $ k $ satisfying $ k\leq y $, $ n $ is an arbitrary integer and $ i $ runs from 1 to 10. From the definition of $ P_{l,i}(t) $, one can show that 
\begin{eqnarray}
\dot{P}_{l,i}(t)=\sum\limits_{j=1}^{10}[{-P_{l,i}(t)(R_{i \to j}^{+} + R_{i \to j}^{-})} \nonumber\\
+ P_{l-1,j}(t)R_{j \to i}^{+} + P_{l+1,j}(t)R_{j \to i}^{-} ] .  \label{dp}
\end{eqnarray}  
By defining
\begin{equation}
f_i(x,t)=\sum\limits_{n=-\infty}^{\infty}{P_{k+5n,i}(t)x^{k+5n}},   \label{defif}
\end{equation}
we immediately see that
\begin{equation}
<Q(t)>=e^*(\frac{d}{dx}{\sum\limits_{i=1}^{10}{f_i}})|_{x=1}
\end{equation}
and
\begin{equation}
<\delta Q^2(t)>={e^*}^2{(\frac{d}{dx}x\frac{d}{dx}{\sum\limits_{i=1}^{10}{f_i}})|_{x=1}}-<Q(t)>^2.
\end{equation}
The time evolution of $ f_i(x,t) $ is governed by the following set of equations which can be derived from Eqs. (\ref{dp}) and (\ref{defif}).
\begin{equation}
\begin{array}{llr}
\frac{d}{dt}f_i(x,t)= & \sum\limits_{j=1}^{10}[{-f_i(x,t)(R_{i \to j}^{+} + R_{i \to j}^{-})} \\
& +xf_j(x,t)R_{j \to i}^{+} + \frac{1}{x}f_j(x,t)R_{j \to i}^{-} ] . \label{kex}
\end{array}
\end{equation}
Eq. (\ref{kex}) is reduced to Eq. (\ref{ke}) when $ x=1 $ and $ f_i(x=1,t)=f_i(t) $  as it is clear from their definitions.
The above set of equations can be rewritten in a vector form $ \dot{\vec{f}}(x,t)=M(x)\vec{f}(x,t) $  where $ M(x) $  is a real $10\times10$ matrix. The solution of Eq. (\ref{kex}) has the form $ f_i(x,t)=\sum\limits_{k=1}^{10}{g_{ik}(x,t){e^{\lambda_k(x)t}}} $,  where $ \lambda_k(x) $ are the eigenvalues of $ M(x) $ and $ g_{ik}(x,t)e^{\lambda_k(x)t} \to 0 $  for $ t \to +\infty $ when $ \lambda_k(x) $ is negative. From Eq. (\ref{kex}), we see that $ M(x) $ has diagonal elements which are negative, off-diagonal elements which are positive or zero, and $ \sum\limits_{j=1}^{10}M_{ij}(x=1)=0 $ . Rohrbach theorem\cite{Fel} tells us that zero is a non-degenerate eigenvalue of $ M(x) $  and all other eigenvalues are negative when $ x=1 $.
If we denote $ \lambda (x) $ as the unique eigenvalue which has the property $ \lambda (x=1)=0 $ then $ \sum\limits_{i=1}^{10}f_i(x,t) $ can be written as the sum of $ g(x)e^{\lambda (x)t} $  and other unimportant terms which will eventually go to zero when we set $ x=1 $ at the end of the calculations at large $ t $. Together with the normalization condition $ g(x=1)=1 $, we can show that 
\begin{equation}
I=\frac{<Q(t)>}{t}= \frac{e^{*} \frac{d}{dx}{(g(x)e^{\lambda(x)t})|_{x=1}}}{t} = e^*{\frac{d\lambda (x)}{dx}}|_{x=1} ,
\end{equation}
\begin{equation}
S=\frac{<\delta Q^2(t)>}{t}={e^*}^2(\frac{d^2\lambda(x)}{dx^2}|_{x=1} + \frac{d\lambda (x)}{dx}|_{x=1}),
\end{equation}
and the Fano factor
\begin{equation}
\tilde{e}={e^*}(1+\frac{d^2\lambda(x)}{dx^2}|_{x=1}/{\frac{d\lambda (x)}{dx}|_{x=1}}).
\end{equation} 
The derivatives $ \frac{d\lambda (x)}{dx}|_{x=1} $ and $ \frac{d^2\lambda(x)}{dx^2}|_{x=1} $ can be determined by applying the first and second derivative operators on the characteristic equation, $ \det[M(x)-\lambda (x)I]=0 $, and setting $ x=1 $ at the end of the calculation.

For the purpose of calculating the first and second derivative of $ \lambda (x) $, we only need to know the lowest order terms in $ \lambda (x) $  of the characteristic equation because of the property $ \lambda (x=1)=0 $. If the characteristic equation is written as:
\begin{equation}
...+A(x){\lambda}^2 + B(x)\lambda + C(x)=0,      \label{chaequ}
\end{equation}
where $ ...$ denotes higher order terms in $ \lambda (x) $, one can show
\begin{equation}
\frac{d\lambda (x)}{dx}|_{x=1}={-\frac{dC(x)}{dx}|_{x=1}}/{\frac{dB(x)}{dx}|_{x=1}}  \label{dlambda}
\end{equation}
and
\begin{equation}
\begin{array}{ll}
\frac{d^2\lambda(x)}{dx^2}|_{x=1}=
\frac{\frac{-d^2C}{dx^2}-2\frac{dB}{dx}\frac{d\lambda (x)}{dx}-2A(x)(\frac{d\lambda (x)}{dx})^2 }{B(x)}|_{x=1} .    \label{d2lambda}
\end{array}
\end{equation}

The kinetic matrix  at zero temperature $ M(x)_{T=0} $ has a simple structure which is shown in Eq. (\ref{tm}). The coefficients $ A(x) $, $B(x)$, $C(x)$ in Eq. (\ref{chaequ}) can be determined from $ M(x)_{T=0}$ after lengthly but straight forward calculations.

\begin{widetext}
\begin{equation}
M(x)_{T=0}=\left(
\begin{array}{cccccccccc}
-P_0 &    0   &    0    &   0   &  0    &     0       &      0      &     0      &    0      &  \frac{xP_4}{\tau^2}  \\
0 &  -(\frac{P_1}{\tau^2}+\frac{P_3}{\tau})  &  0  &    0   &        0    &       0       &      0      &     0      &    xP_3      &  \frac{xP_1}{\tau}    \\
0 &    \frac{xP_1}{\tau^2}    &    -P_2   &     0    &        0    &       0       &      0      &     0      &    0      & 0             \\
xP_0 &    \frac{xP_3}{\tau}   &     0    &  -(\frac{P_0}{\tau}+\frac{P_3}{\tau^2} )    &        0    &       0       &      0      &     0      &    0      & 0             \\
0 &    0   &     0    &     \frac{xP_3}{\tau^2}     &     -P_4    &       0       &      0      &     0      &    0      & 0             \\
0 &    0   &     xP_2   &     \frac{xP_0}{\tau}    &        0    &    - (\frac{P_0}{\tau^2}+\frac{P_2}{\tau} )    &        0      &     0      &    0      & 0             \\
0 &    0   &     0    &     0    &       0    &       \frac{xP_0}{\tau^2}      &    -P_1     &     0         &    0        & 0             \\
0 &    0   &     0    &     0    &        xP_4    &       \frac{xP_2}{\tau}       &      0      &   -(\frac{P_2}{\tau^2} +\frac{P_4}{\tau} )   &    0      & 0             \\
0 &    0   &     0    &     0    &        0    &       0       &      0      &     \frac{xP_2}{\tau^2}      &   -P_3   & 0             \\
0 &    0   &     0    &     0    &        0    &       0       &      xP_1     &     \frac{xP_4}{\tau}     &    0      & -(\frac{P_4}{\tau^2}+\frac{P_1}{\tau}  )        \\        \label{tm}
\end{array}
\right)
\end{equation}
\end{widetext}

\begin{figure}
\includegraphics[width=2.8in]{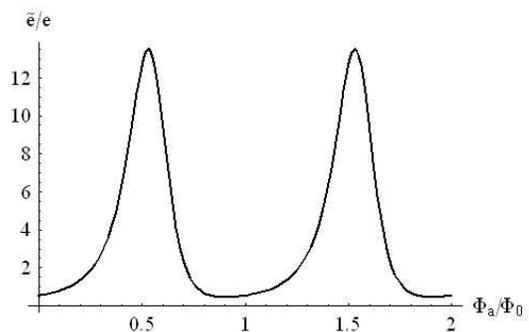}
\caption{\label{Fano} A plot of the Fano factor $ \tilde{e}(\Gamma_1,\Gamma_2)/e $ against $ \Phi_{a}/\Phi_{0} $ with $ \Gamma_1 = \Gamma_2 =1 $.}
\end{figure}

The flux dependence of the Fano factor $ \tilde{e} $ with $ \Gamma_{1}=\Gamma_{2}=1 $ is depicted in Fig. \ref{Fano}. We can see that when the current is minimal at $ \Phi_a/\Phi_0=1/2 $, $ \tilde{e} \approx 12.7e $, which is much larger than the maximum Fano factor in the case of both the $ \nu=1/5 $ Laughlin state with $ \tilde{e}_{max} =1 $ and the Moore-Read state with $ \tilde{e}_{max} \approx 3.2 e $. The large Fano factor can be understood in the following way: suppose we tune the applied magnetic field such that $ \Phi_{a}/\Phi_{0} \approx 1/2 $ and $ p_{2} \ll 1 $. The system can be trapped in state $ 3 $ for a long time before a tunneling event can happen which would drive the system to state $ 6 $. Once the system is in state $ 6 $, a series of tunneling events can happen in a relatively short period of time which would drive the system to other states before getting trapped at state $ 3 $ again. As a result, the Fano factor can be much larger than 1 electron charge. Since the existence of those by-pass roads (by-passing state $ 3 $) is a direct consequence of the non-trivial fusion rules and braiding rules, the observation of a Fano-factor which is larger than one electron charge would provide evidence for the existence of non-Abelian statistics in FQH states.

All the calculations above were performed in the case of $ k=3 $ Parafermion state, which corresponds to a FQH state with $ \nu=13/5 $. As we mentioned before, the $ \nu=12/5 $ state can be obtained by a particle-hole transformation,\cite{RR,CS} which results in a change in a sign of statistical angles as well as a sign change in the charge carried by the tunneling quasiparticle. Consequently, the tunneling probabilities $ p_l $ in Eq. (\ref{tp}) will be changed to $ p'_l=p_{-l} $. The net effect is simply the same as changing $ \Phi_a $ to $ -\Phi_a $ and $ \delta $ to $ -\delta $ at the end of the calculations which will not affect our arguments. 

\section{conclusions}

In conclusion, we have shown that in Mach-Zehnder interferometer geometry, the tunneling current and the zero temperature shot noise can be used to show the existence of $ k=3 $ parafermion statistics in the $ \nu=12/5 $ FQH state. More specifically, the scaling exponent $ b $ in the relation between the flux dependent part and the flux independent part of the tunneling current in Eq. (\ref{ep}) is $ 5/2 $ in the $ \nu=12/5 $ Read-Rezayi state, in contrast with $ b=2 $ in the Moore-Read state, but is the same as the case in the $ \nu=1/5 $ Laughlin state. However, I-V curve in the $\nu=12/5 $ Read-Rezayi state is asymmetric. This property is absent in all Laughlin states. In addition, the Fano factor in the $\nu=12/5 $ Read-Rezayi state can exceed $ 12.7e $ when the current is tuned to its minimum value by changing the applied magnetic flux. This number well exceeds the maximum possible Fano factor in all Laughlin states and the $ \nu=5/2 $ Moore-Read state which was shown to be $ e $ and $ 3.2 e $ respectively. The two key properties, namely, the asymmetric I-V curve and the larger than one electron charge Fano factor, are direct consequences of non-Abelian statistics. Their experimental observations would provide evidence for the existence of non-Abelian statistics in the $ \nu=12/5 $ FQH state.

\begin{acknowledgments}
The author is indebted to D. Feldman for his encouragement and guidance, as well as inspiring discussions throughout this project. Useful discussions with A. Kitaev, J.B. Marston, F. Sch\"{u}tz, A. Stern and X.G. Wen are also gratefully acknowledged. This work was supported by the National Science Foundation under Grant No. DMR-0544116.
\end{acknowledgments}

\end{document}